# EVOLUTION OF THE DISORIENTED CHIRAL CONDENSATES IN THE MEAN FIELD APPROXIMATION

YUVAL KLUGER*

*Theoretical Division, Los Alamos National Laboratory, Los Alamos, New Mexico 87545, USA*

ABSTRACT

We study the dynamics of the chiral phase transition expected during the expansion of the quark-gluon plasma produced in a high energy hadron or heavy ion collision, using the $O(4)$ linear sigma model in the mean field approximation. Starting from an approximate equilibrium configuration at an initial proper time $\tau$ in the disordered phase, we study the transition to the ordered broken symmetry phase as the system expands and cools.

We give results for the proper time evolution of the effective pion mass and for the pion two point correlation function. We investigate the possibility of disoriented chiral condensate being formed during the expansion. In order to create large domains of disoriented chiral condensates low-momentum instabilities have to last for long enough periods of time. Our simulations show no instabilities for an initial thermal configuration. For the far-of-equilibrium cases studied, the instabilities are formed during the initial stages of the expansion survive for short proper times. For slow expansion rates even such configurations do not develop instabilities.

---

## 1. Model and Approximations

The possibility of producing large correlated regions of quark condensate $<\bar{q}_i q_j>$ pointing along the wrong direction in isospin space was proposed [1-2] to explain rare events where there is a deficit or excess of neutral pions observed in cosmic ray experiments[3]. The idea that such disoriented chiral condensates (DCC's) can be formed has been the source of several experimental proposals[4] in high energy collisions. It has been recognized that nonequilibrium dynamics can yield regions of DCC[5-10]. These authors introduced instabilities of low momenta modes by quenching the system, i.e., evolving a "hot" state under the dynamics of a zero temperature hamiltonian.

There are clearly two important questions that the theoretical models should aim to answer. The first one is to determine whether during the evolution that follows the collision there are instabilities affecting the fluctuations, in which case there is a chance of the correlations growing.

The second question is, assuming the instability exists, can the correlated domains grow large enough so that many pions can be emitted from each domain making the detection of DCC's possible. Experimentally, one can look at the probability distribution of the fraction of neutral pions $P(f)$ where $f = n_{\pi^0}/(n_{\pi^0} + n_{\pi^\pm})$. Assuming that all isospin orientations of $\vec{\pi}$ are equally distributed so that $P(\Omega) = const$, and using the equation $P(\Omega)d\Omega = P(f)df$, one finds that for an ideal case where only one large domain is formed this distribution is given by $P(f) = 1/2\sqrt{f}$, provided that $n^i \propto |\pi^i|^2$, where $f = \cos^2\theta$. If the final particle distribution results from uncorrelated events, one would expect the neutral fraction to be distributed binomially around $f = 1/3$. Thus, for an ensemble with domains of different sizes the resulting distribution $P(f)$ dwells somewhere in between binomial and domain-like distribution. To address these questions we employ the $O(4)$ linear sigma model, in the mean field approximations, which seems to have the essential attributes of being simple but realistic enough: it has the correct chiral symmetry properties and also appropriately describes the low energy phenomenology of pions. Moreover, the mean field approximation improves the classical approximation, because it takes into account quantum fluctuations that can not be neglected when the system cools down. In this model the mesons are organized in an $O(4)$ vector $(\phi_1, \phi_2, \phi_3, \phi_4) = (\sigma, \pi_1, \pi_2, \pi_3)$. The effective action is derived by using the the large $N$ expansion of a $O(N)$ model, and is given by

$$\Gamma[\bar{\Phi}, \bar{\chi}] = \int d^4x \{-\frac{1}{2}\sum_{i=1}^N \bar{\phi}_i(\Box + \bar{\chi})\bar{\phi}_i + \frac{\bar{\chi}^2}{4\lambda} + \frac{1}{2}\bar{\chi}v^2 + H\bar{\sigma} + \frac{i}{2}N\,\mathrm{tr}\,\ln G_0^{-1}\}, \quad (1)$$

where $G_0^{-1}(x,y) = i[\Box + \bar{\chi}(x)]\,\delta^4(x-y)$, and $\chi = \lambda(\sum_{i=1}^N \phi_i\phi_i - v^2)$ with $N = 4$. Varying the effective action with respect to the mean fields, we derive their equations of motion

$$(\Box_x + \bar{\chi}(x))\bar{\phi}_i(x) = H\delta_{i1} \quad (2)$$

$$\bar{\chi}(x) = \lambda(-v^2 + \sum_{i=1}^{N}[\bar{\phi}_i{}^2(x) + G_0(x,x)]). \tag{3}$$

The functions $G_0(x,x)$ that appear in (3) are the coincidence limit of the propagators $G_0(x,y)$ that invert the operator $G_0^{-1}$.

By calculating two and four point function the parameters $\lambda$, $H$ and $v$ were chosen to give the best fit to the physical observables $f_\pi$, $m_\pi$ and the $\delta_s^{I=0}$ phase shifts[10].

In Bjorken's Baked Alaska picture boost invariant initial conditions are fixed at an initial proper time $\tau_0$ (these initial configurations eventually lead to a flat rapidity distribution of pions as expected in the future RHIC experiments). Dynamics then incorporates a cooling mechanism that may lead to low momentum instabilities (in contrast to a crude quench approximation that plants the instabilities from the beginning). In the boost invariant picture it is natural to use the following set of coordinates

$$\tau \equiv (t^2 - z^2)^{1/2}, \qquad \eta \equiv \frac{1}{2}\log(\frac{t+z}{t-z}), \qquad \mathbf{x}_\perp, \tag{4}$$

where $\tau$ is the proper time and $\eta$ is the spatial rapidity. The equations of motion reduce to

$$\tau^{-1}\partial_\tau \tau \partial_\tau \bar{\phi}_i(\tau) + \bar{\chi}(\tau) \bar{\phi}_i(\tau) = H\delta_{i1}, \tag{5}$$

$$\bar{\chi}(\tau) = \lambda(-v^2 + \sum_{i=1}^{N}[\bar{\phi}_i{}^2(\tau) + <\varphi^2(\eta, x_\perp, \tau)>]), \tag{6}$$

$$\left(\tau^{-1}\partial_\tau \tau \partial_\tau - \tau^{-2}\partial_\eta^2 - \partial_\perp^2 + \bar{\chi}(\tau)\right)\varphi(\eta, x_\perp, \tau) = 0, \tag{7}$$

where we have introduced an auxiliary field $\varphi(x,\tau)$, related to the Wightman function by:

$$G_0(\eta_x, x_\perp, \tau_x, \eta_y, y_\perp, \tau_y) \equiv <\varphi(\eta_x, x_\perp, \tau_x)\varphi(\eta_y, y_\perp, \tau_y)>. \tag{8}$$

At equal proper times

$$G_0(\eta_x, x_\perp, \eta_y, y_\perp; \tau) = G_c(\eta_x, x_\perp, \eta_y, y_\perp; \tau), \tag{9}$$

where $G_c(x,y) \equiv \frac{1}{N}\sum_{i=1}^{N} <\phi_i(x)\phi_i(y)> - \bar{\phi}_i(x)\bar{\phi}_i(y)$. We expand the field $\varphi$ in an orthonormal basis

$$\varphi(\eta, x_\perp, \tau) \equiv \frac{1}{\tau^{1/2}}\int dk(\exp(ik\cdot \mathbf{x})f_k(\tau)\, a_k + h.c.), \tag{10}$$

where $k\cdot \mathbf{x} \equiv k_\eta \eta + \vec{k}_\perp \vec{x}_\perp$, $dk \equiv dk_\eta d^2k_\perp/(2\pi)^3$, and the mode functions $f_k(\tau)$ evolve according to:

$$\ddot{f}_k = -(\frac{k_\eta^2}{\tau^2} + \vec{k}_\perp^2 + \bar{\chi}(\tau) + \frac{1}{4\tau^2})f_k \equiv -\omega_k^2 f_k. \tag{11}$$

A dot here denotes the derivative with respect to the proper time $\tau$. The Fourier transform of the expectation value $<(\varphi)^2(x,\tau)>$ can be expressed in terms of the mode functions $f_k$ and of the distribution functions

$$n_k \equiv <a_k^\dagger a_k>, \quad g_k \equiv <a_k a_k>, \quad g_k^* \equiv <a_k^\dagger a_k^\dagger> \tag{12}$$

which entirely characterize the initial state of the quantum field.

In order to solve equations (5), (6) and (11) as an initial value problem, we need to fix at $\tau_0$ the mean values $\bar{\phi}_{ik}$, $\dot{\bar{\phi}}_{ik}$, $n_k$, $g_k$ and $g_k^*$. This amounts to specifying five initial conditions at each momentum $k$. * In the boost invariant case $\bar{\phi}_{ik}(\tau) = \dot{\bar{\phi}}_{ik}(\tau) = 0$ for $k \neq 0$.

The complexity of the processes we study suggests that our initial configuration is not a pure state but a mixed ensemble that should be described by a density matrix. In the mean field approximation the most general density matrix for each mode $k$ is a gaussian with five parameters associated with the above initial conditions and it can be written as

$$\rho(\phi_{ik}, \phi'_{ik}) = \frac{1}{\sqrt{2\pi G_{ik}}} \exp\{i\dot{\bar{\phi}}_{ik}(\phi'^{*}_{ik} - \phi^*_{ik}) - \frac{1}{2}(K_{ik} - 4G_{ik}\Sigma^2_{ik})|\phi'_{ik} - \phi_{ik}|^2$$
$$-i\Sigma_{ik}[|\phi_{ik} - \bar{\phi}_{ik}|^2 - |\phi'_{ik} - \bar{\phi}_{ik}|^2] - \frac{1}{8G_{ik}}|\phi_{ik} + \phi'_{ik} - 2\bar{\phi}_{ik}|^2\}, \tag{13}$$

where

$$G_{ik} = <\phi_{ik}\phi^*_{ik}> - \bar{\phi}_{ik}\bar{\phi}^*_{ik} \tag{14}$$

$$K_{ik} = <\dot{\phi}_{ik}\dot{\phi}^*_{ik}> - \dot{\bar{\phi}}_{ik}\dot{\bar{\phi}}^*_{ik} \tag{15}$$

$$4\Sigma_{ik}G_{ik} = <\{\phi_{ik}, \dot{\phi}^*_{ik}\}> - \{\bar{\phi}_{ik}, \dot{\bar{\phi}}^*_{ik}\}. \tag{16}$$

Pure density matrix satisfies the condition $p \equiv K - 4G\Sigma^2 + 1/4G = 0$ (no $\phi_i(k)\phi'_i(k)$ mixed term). This subset corresponds to squeezed-coherent states[11]. For $\Sigma = 0$ and $G_{ik} = 1/\sqrt{2\omega_k}$, the density matrix operator becomes $\hat{\rho}^c = |c><c|$ where $|c>$ is a coherent state. One can make a direct connection between our calculation and earlier work[5-8], which used classical approximation to the same model. The equilibrium density matrix of a free theory is Gaussian and in the high $T$ limit ($\beta\omega_k << 1$) has the following representation:

$$\rho = \prod_{ik} \int d\bar{\phi}_{ik}d\dot{\bar{\phi}}_{ik} \exp\{-\frac{\beta}{2}(|\dot{\bar{\phi}}_{ik}|^2 + (k^2 + m^2)|\bar{\phi}_{ik}|^2)\}\rho^c_{ik}(\bar{\phi}_{ik}, \dot{\bar{\phi}}_{ik}). \tag{17}$$

Therefore, solving the mean field equations with an initial thermal homogeneous density matrix is equivalent to solving the semi classical equations for a thermal ensemble of inhomogeneous "classical" initial conditions. We want to stress the point that in the framework of the the mean field approximation one has to fix five initial conditions to describe the whole ensemble, instead of separately calculating the equations of motion for each member of the ensemble, as has been done in[5-8].

---

*The formulation of Eqs. (5), (6) and (11) may give an impression, that we have to specify extra four initial conditions for $f_k(\tau_0)$ and $\dot{f}_k(\tau_0)$. This impression is false, since as it can be easily checked that the change of $f_k(\tau_0)$ and $\dot{f}_k(\tau_0)$ if accompanied by an appropriate change of $n_k$ and $g_k^*$ according to a Bogolyubov transformation leaves $G_c$, $\dot{G}_c$ and $\ddot{G}_c$ invariant, and these quantities (together with $\bar{\phi}_{ik}$, $\dot{\bar{\phi}}_{ik}$) unambiguously specify the initial data. Therefore, we will fix the initial data $f_k(\tau_0)$ and $\dot{f}_k(\tau_0)$ so that the vacuum state coincides with the ordinary Minkowsky vacuum, at least for high momentum.

In the interacting case, $\hat{\rho}^{th} = e^{-\beta H}$ is not a gaussian any more, and therefore it is not contained in the set of mean field density matrices. However, one can approximate the thermal density matrix by replacing the quartic terms by using the thermal mass. We note that for a gaussian density matrix the requirement of isospin invariance leads immediately to a factorization form $\rho = \prod_i \rho_i$. Due to the factorizability of the density matrix, if one species $i$ has a large average $< N^i >$, the whole density matrix will also have a large $< N >$ and vice versa. This is also true for the variance $< (N - \bar{N})^2 >$. The mean field approximation can only accommodate ensembles with either large fluctuations in both the total number of pions and number of neutral pions or small fluctuations in both observables. Ensembles that have sharply defined total number of particles but large fluctuations in the number of neutral pions can not be described in this framework.

## 2. Results

In Fig. 1 we display the time evolution of the effective mass $\bar{\chi}$, the mean density of particles $< n > \equiv < dN/d\eta dx_\perp >$ and the relative fluctuation of $dN/d\eta dx_\perp$ starting at initial proper time $\tau_0 = 1 fm/c$. We also plot the time evolution of the fourier transform of the correlator $G_c(\Delta \eta, \Delta x_\perp; \tau)$ at the planes $|\mathbf{k}_\perp| = 0$ and $k_\eta = 0$. The left row corresponds to an initial thermal configuration. In this case $\bar{\chi}$ does not become negative during the evolution so that no instabilities develop. The density $< dN/d\eta dx_\perp >$ grows very mildly and so does its relative fluctuation. The fluctuation at large proper times is almost equivalent to its initial value. Therefore, observing a DCC for an initial thermal ensemble is unlikely. From the inverse width at half height of the profile of the Fourier transform we learn that the size of a domain in the beam direction is about one unit of space-time rapidity and about $1 fm$ or less in the transverse direction. In the right row we generated instabilities by choosing an initial ensemble that is far from equilibrium for which we modified only the initial condition for $\dot{\bar{\sigma}}$ from 0 to $-1$. We see that $\bar{\chi}$ becomes negative at early stages of the evolution and at this time an exponential growth of the low momenta modes occur. The density $< dN/d\eta dx_\perp >$ triples itself by sucking energy from the $\bar{\sigma}$ field and reaches a plateau after few tens of $fermi/c$. The relative fluctuation of this density also grows substantially and therefore the observation of DCC is more likely for this "peculiar" ensemble. This is also reflected in the structures appearing in the fourier transforms of $G_c$. More specifically we see that at the transverse direction there is an amplification of the amplitude at low momenta, which corresponds to a domain of about $2 fm$. However, in this case it is hard to define the size of the domain by naively measuring the inverse width due to multiple scales that emerge during the dynamics. For similar out of equilibrium ensembles we found no instabilities if we start the simulations at proper times greater than $2fm/c$. This implies that if the expansion rate is too slow or if we start with lower energy densities even for out of equilibrium ensembles, DCC will be hardly produced.

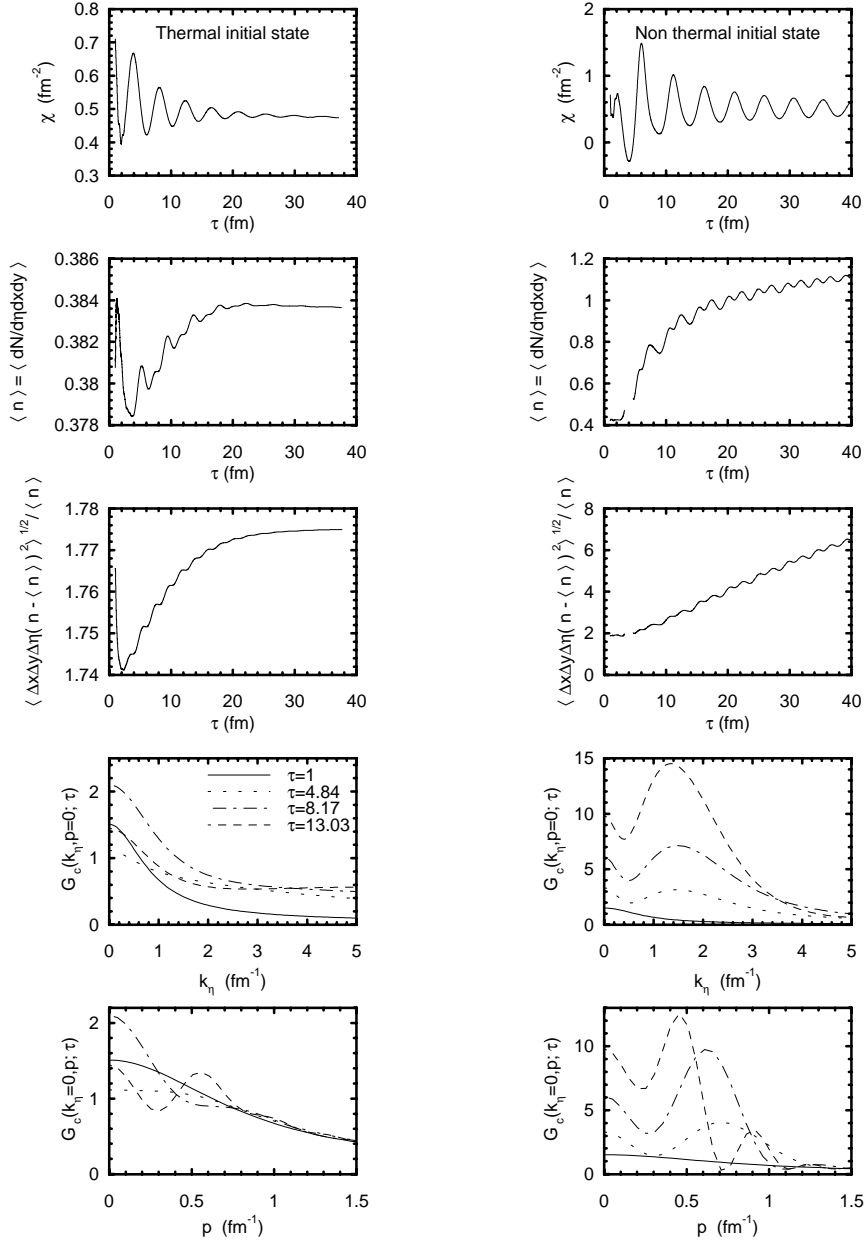

Fig. 1. Proper time evolution of the $\bar{\chi}$ field, $< dN/d\eta dx_\perp >$, the relative fluctuation of $dN/d\eta dx_\perp$ times the square root of the extent of the system $(\Delta x_\perp \Delta \eta)^{1/2}$ and of the fourier transforms of the Green's function $G_c$ at the planes of $p \equiv |\mathbf{k}_\perp| = 0$ and $k_\eta = 0$. The left row is for an initial thermal distribution where $n_k^i = 1/(\exp \beta \omega_k - 1)$, $\bar{\sigma}(\tau_0 = 1) = \bar{\sigma}_{eq}(T = 200 MeV)$, $\bar{\pi}^i(1) = \dot{\bar{\pi}}^i(1) = \dot{\bar{\sigma}}(1) = 0$. The right row has the same initial conditions as in the left one, but for $\dot{\bar{\sigma}}(1) = -1$.